\documentclass[conference]{IEEEtran}
\IEEEoverridecommandlockouts

\usepackage{cite}
\usepackage{amsmath,amssymb,amsfonts}
\usepackage{algorithmic}
\usepackage{graphicx}
\usepackage{textcomp}
\usepackage{xcolor}

\begin{document}

\title{Parallel Machine Learning of Partial Differential Equations}

\author{\IEEEauthorblockN{Amin Totounferoush\IEEEauthorrefmark{1},
		Neda Ebrahimi Pour\IEEEauthorrefmark{2},
		Sabine Roller\IEEEauthorrefmark{2} and
		Miriam Mehl\IEEEauthorrefmark{1}}
	\IEEEauthorblockA{\IEEEauthorrefmark{1}Institute for Parallel and Distributed Systems\\
		University of Stuttgart,
		Germany\\ Email: \{amin.totounferoush, miriam.mehl\}@ipvs.uni-stuttgart.de}
	\IEEEauthorblockA{\IEEEauthorrefmark{2}Chair of Simulation Techniques and Scientific Computing\\
		University of Siegen,
		Germany \\Email:\{neda.epour, sabine.roller\}@uni-siegen.de }}

\maketitle

\begin{abstract}
In this work\footnote{preprint submitted to PDSEC 2021, IPDPS Conefernce.}, we present a parallel scheme for machine learning of partial differential equations. The scheme is based on the decomposition of the training data corresponding to spatial subdomains, where an individual neural network is assigned to each data subset. Message Passing Interface (MPI) is used for parallelization and data communication. We use convolutional neural network layers (CNN) to account for spatial connectivity. We showcase the learning of the linearized Euler equations to assess the accuracy of the predictions and the efficiency of the proposed scheme. These equations are of particular interest for aeroacoustic problems. A first investigation demonstrated a very good agreement of the predicted results with the simulation results. In addition, we observe an excellent reduction of the training time compared to the sequential version, providing an almost perfect scalability up to 64 CPU cores.          
\end{abstract}

\begin{IEEEkeywords}
parallel training, convolutional neural networks, partial differential equation 
\end{IEEEkeywords}

\section{Introduction}
	
Machine learning methods have shown promising performance in various scientific fields, ranging from computer vision to disease diagnosis and personalized therapy. One potential usage of these methods is the modeling of the behavior of a physical system, without using the classical mathematical equations. The classical models frequently suffer from very costly solution processes. A data-driven modeling approach has the capability of resolving such issues. Many studies have been recently conducted in this area, covering a wide range of problems and purposes, such as equation solution, parameter analysis or uncertainty quantification. For example, Vlachas et al.~\cite{vlachas2018data} extend the long short-term memory (LSTM) networks for data driven forecasting. The weather forecasting can be named as a potential application of this work. Raissi et al.~\cite{raissi2019physics} introduced the physics-informed neural networks. These networks add the governing equations to the loss function to force the training process to obey the physical laws. These methods are more effective when only a small amount of data is available.  

Given the size of each data set in typical simulation applications, efficient training and inference becomes very important. For instance, training a neural network for weather prediction can be very costly due to the large amount of measurement points, both spatially and temporally. There are various approaches in literature to improve the efficiency of the training, i.e., to reduce the time-to-train, while preserving the learning quality. These approaches can be generally divided into data parallelization and model parallelization schemes~\cite{ben2019demystifying}. The first approach divides the data among different processes while the second approach shares all data among processes but distributes the computation among processes. Both approaches require data communication for synchronization. For instance, Vivian et al.~\cite{viviani2019deep} presented a data parallel approach, where the available training data are split into smaller chunks. Each chunk is given to a network and one step training is applied. Through a global reduction operation, the networks resulting from different training data chunks can share their weights. The weights are averaged and constitute a new network, which is shared among all individual MPI ranks. This procedure is repeated until all available data are fed into the network and the training completed. This approach is able to reduce the training time. However, it alters the learning algorithm resulting in decreased learning. In addition, the global reduction operations are potential performance bottlenecks. 

In the current study, we propose a novel parallelization scheme for the training of deep neural networks (DNNs). The proposed method is primarily targeting simulations in different scientific areas, but can be generalized to be utilized for other fields as well. For simulation applications, DNNs can be used to either replace classical physics-based models (data-driven approach) or to enhance or dynamically adapt them (data-integrated approach). To optimize the time-efficiency of the training and, thus, maximize the benefit of such an approach, we propose to decompose the individual training data sets into smaller spatial sections, each, instead of distributing complete data sets among processes. Thus, each network learns the data for its subdomain and the training phase is communication-free. For inference, the method requires only boundary data exchange, which can be done in a parallel point-to-point way similar to classical simulations.
 
The rest of the paper is organized as follows: In Sec.~\ref{sec:neural}, we present the architecture of the neural network, used for each subdomain. Sec.~\ref{sec:parallel} explains the MPI parallelization of the training and the inference. In Sec.~\ref{sec:results}, we showcase an aeroacoustics problem, where the linearized Euler equations are learned. We discuss both the accuracy of prediction and the scalability of the presented scheme. Finally, Sec.~\ref{sec:conclusion} summarizes and concludes the paper.

\section{Architecture of Neural Network}
\label{sec:neural}
\begin{figure*}[!hbpt]
	\centering
	\includegraphics[width=\textwidth]{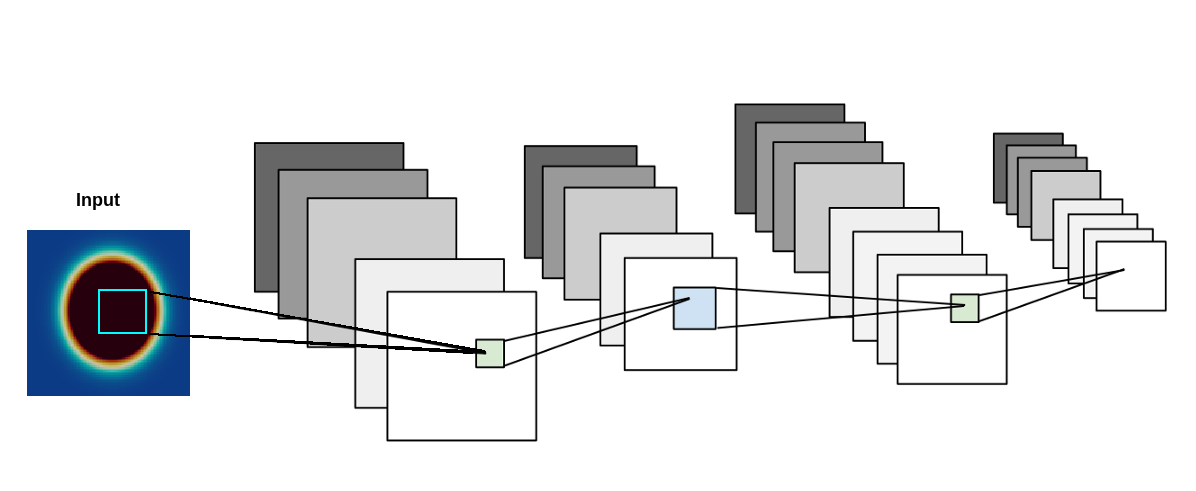}
	\caption{A multi-layer and multi-channel convolutional neural network. The input consists of domain variables measured at various data points in the domain.}
	\label{fig:cnn}
\end{figure*}

The current work aims to improve the efficiency of DNNs in simulation science. To preserve the spatial connectivity that exists in these sort of applications, we use convolutional (CNN) layers within the network. However, the proposed parallelization scheme can be incorporated with other type of layers. CNN layers incorporate a kernel with learnable elements to map a multi-dimensional input to an output (see Fig.~\ref{fig:cnn}). This can be used for both regression and classification, depending on the final layer of the network. In the rest of this section, we summarize the configuration of the neural network used in the current work. \vspace{1mm}

\noindent\textbf{Layers:} We use four CNN layers with the following input and output channels presented in Table~\ref{tb:layers}.

\begin{table}[!hbpt]
	\centering
	\caption{\label{tb:layers}CNN layers architecture: Output channels of one layer should match with the input channels of the next layer.}
	\begin{center}
		\begin{tabular}{c|c|c|c|c} \hline 
			layer& input & output & kernel & Padding \\
			number & channels & channels& size &  \\
			\hline
			1 & 4 & 6 & $4\times6\times5\times5$ & Yes\\
			\hline
			2 & 6 & 16 & $6\times16\times5\times5$ & Yes\\
			\hline
			3 & 16 & 6 & $16\times6\times5\times5$ & Yes\\
			\hline
			4 & 6 & 4 & $6\times4\times5\times5$ & Yes\\
			\hline	
		\end{tabular}
		%		}
	\end{center}
\end{table}

As we intend to mimic the solution of the two-dimensional linearized Euler equations, the network input and output must have four channels, corresponding to the training data types: pressure, density, velocity in x-direction and velocity in y-direction. 
 
\noindent\textbf{Activation function:} To define a neural network, we need to select a non-linear activation function $\sigma$, which transforms the input of a neuron to its output. There are plenty of suggestions for possible functions with different properties and use cases. Glorot et al.~\cite{rect} showed that rectified linear units (ReLUs) have a better performance for training the NNs than sigmoid and hyperbolic tangent functions. Since 2017, they are therefore the most commonly used activation function in deep learning ~\cite{ramachandran2017searching}. In the simplest case, $\sigma$ is given by

\begin{equation}
\label{eq:rect}
\sigma(x) = \max(0, x).
\end{equation}

Although, the gradient is zero for all negative inputs, it does not vanish for large $x$ (as it does for sigmoid activation, e.g.). A vanishing gradient stops the network's learning and weights are not updated any more. At $x=0$, the gradient is undefined, but in practice, $x=0$ very rarely occurs. Nonetheless, a value for this unlikely case should be selected, e.g., zero. The problem for negative inputs can be fixed with leaky ReLUs

\begin{equation}
\label{eq:leakyrect}
\sigma(x) =
\begin{cases}
x & \text{for } x \geq 0, \\
\varepsilon x & \text{for } x < 0,
\end{cases}
\end{equation}

where $\varepsilon$ is some small value such as $\varepsilon = 0.01$. Alternatively, $\varepsilon$ can be considered as a parameter that must be learned alongside the weights. We use a constant $\varepsilon = 0.01$ in the current work. 

\noindent\textbf{Optimizer:} The most important factor for creating well-performing neural networks are the entries of the weight matrices $W$. During training, we essentially solve an optimization problem minimizing the loss for the given training data. The choice of the optimizer is heavily contributing to the result. All common optimizers work in the same fashion:
They begin with some initial parameter estimate $W^{(0)}$ and update this estimate according to some rule $W^{(t + 1)} \gets W^{(t)} + \eta^{(t)} \vec{s}^{(t)}$.
Their main difference is the choice of the search-direction $\vec{s}$ and the learning-rate (or step-size) $\eta^{(t)}$ for step $(t)$. The parameters are updated until either a maximum number of iterations is reached or some convergence criterion is fulfilled. After trying different available options, we found the ADAM \cite{adam} optimizer to have the best performance in our case.

ADAM is based on the well-known stochastic gradient descent method (SGD) \cite{Goodfellow-et-al-2016}, but it uses a concept called \emph{momentum} to further improve the convergence of the optimization. The so-called momentum is conceived by observing a common problem in Gradient descent based optimizers. In cases where the eigenvalues of the Jacobian of the loss function with respect to  the weights vary strongly, gradient descent oscillates between slopes while only slowly converging towards the minimum \cite{qian1999momentum}. Momentum ($m^{(t)}$) aims to improve the convergence by adding a fraction $\rho_{1} \in [0, 1)$ of the the previous search direction to the current one

\begin{equation}
\label{eq:firstmoment}
\begin{split}
& m^{(t)}  := \rho_{1} m^{(t-1)} + (1-\rho_{1}) \frac{d L}{d W}\\
& W^{(t + 1)} \gets W^{(t)} - m^{(t)},
\end{split}
\end{equation}

where $m^{(t)}$ is the momentum, $L$ is the loss function and $W$ are weight matrix entries. This speeds up the optimization towards the bottom of the ravine and thus the minimum. In addition to the first momentum in equation (\ref{eq:firstmoment}), ADAM uses the second momentum

\begin{equation}
\label{eq:secondmoment}
v^{(t)} := \rho_{2} v^{(t - 1)} + (1- \rho_{2}) \frac{d L}{d W} \odot \frac{d L}{d W},
\end{equation}

where $\odot$ is the entry-wise product. The moments are initially set to be zero-vectors $m^{(0)} = \vec{0},~ v^{(0)} = \vec{0}$, which leads to a bias towards \textbf{$\vec{0}$}, in particular during the first few steps. The parameters $\rho_{1}, \rho_{2} \in [0,1)$ counteract this effect by exponential decay adjusting of the moments. 

\begin{equation}
\begin{split}
& \hat{m}^{(t)} := \frac{1}{1 - \rho_{1}^{t}} m^{(t)}, \\
& \hat{v}^{(t)} := \frac{1}{1 - \rho_{2}^{t}} v^{(t)}.
\end{split}
\end{equation}

$\rho_{1}^{t}$ is the $t$-th power of $\rho_{1}$ (the same holds for $\rho_{2}^{t}$). The ADAM update rule is then

\begin{equation}
W^{(t)} \gets W^{(t-1)} - \eta \frac{\hat{m}}{\sqrt{\hat{v} + \epsilon}}.
\end{equation}

Kingma et al. \cite{adam} suggest a global learning-rate of $\eta = 0.01$ and a smoothing value of $\epsilon = 10^{-8}$. Empirical evidence in ~\cite{adam} shows that ADAM outperforms other optimizers in terms of convergence speed and quality of the found solution in many cases.

\noindent\textbf{Loss function:} The most commonly used loss function for neural networks is the mean squared error (MSE). We have observed that the \textbf{mean absolute percentage error} (MAPE) is better suited for our specific application. This is due to the fact that the measured values have different orders of magnitudes and the mean squared error penalizes deviations on the larger data points much more. The MAPE is given by
\begin{equation}
\label{eq:mape}
L^{\text{MAPE}} := \frac{100\%}{m}\sum_{k = 1}^{m} \left| \frac{y_{prediction} - y_{target}}{y_{target}} \right|,
\end{equation}
where $y_{target}$ are the target values and $y_{prediction}$ are the corresponding predictions. Compared to the mean squared error, the MAPE loss is proportional to the magnitude of the data point and the prediction.

\section{Training and Inference Parallelization}
\label{sec:parallel}
We present a novel parallel scheme for efficient and scalable training and inference of neural networks. The goal is to efficiently exploit multi-core machines (CPU or GPU) in order to reduce the training and inference time while preserving the learning quality.

\begin{figure*}[!htbp]
	\centering
	\includegraphics[width=\textwidth]{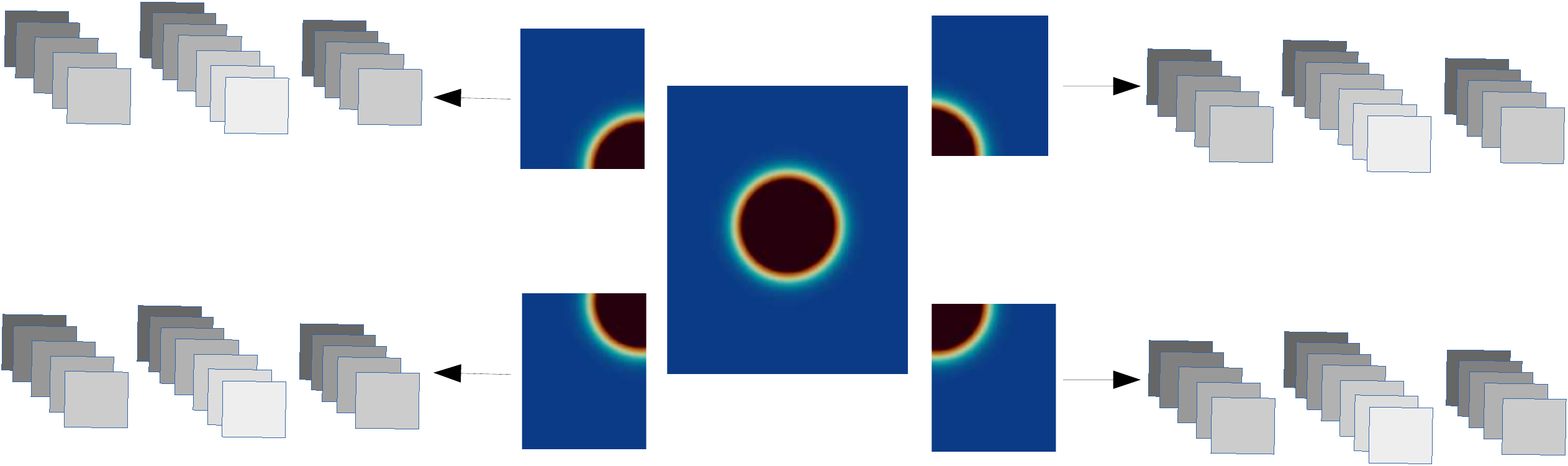}
	\caption{Parallel training of neural networks: The data are decomposed into smaller sections and each section is fed into an individual neural network.}
	\label{parallel-training}
\end{figure*}

\vspace{5mm}
\noindent \textbf{Training:}
We decompose each individual training data set into smaller sections and feed each subsection into an independent neural network. Suppose that we have a set of 1000 training data sets, each representing a two-dimensional square domain with 100 data points. We propose following steps for training:
\vspace{2mm}
\begin{enumerate}
\item Split each data set into smaller sections, for example 4 smaller subdomains, each with only 25 data points.
\item Feed each section into an individual network (see Fig.~\ref{parallel-training}).
\item Use MPI to parallelize the training by assigning an MPI rank to each network. 
\item Use an individual cost function and optimization process for each network.
\item Train each network for the specific part of the domain and use it to predict only its own section.  
\end{enumerate}

Since applying CNNs reduces the size of the two-dimensional input data set by (k-1) rows and (k-1) columns, if we use a $k \times k$ convolution kernel, the network output can not be directly compared to the target data (whereas the input and target data have the same dimensions). For the first layer, we increase the input dimension to match its output dimensions with the target data. Meaning, that input data for neighboring processes are overlapping. This helps both to match dimensions and preserve the spatial connectivity between neighbouring processes. If only a single CNN layer is used, using the larger input removes the mismatch of output and target data dimension. For more than one layer, the following approaches would be possible to solve the dimension mismatch issue:
\begin{enumerate}
	\item Padding the input data with zeros, to achieve the desired size of the output,
	\item Padding the input data with data from neighboring subdomains,
	\item Comparing only the inner $(N-k+1) \times (N-k+1)$ data points of the target data to the network output,
	\item Adding de-convolutional layers or the transpose convolution.
\end{enumerate}

Comparing the inner data points would limit the usability of the output data, as substitutes of the actual simulation (data at subdomain interfaces are missing). Therefore, we currently use only the first two approaches. Applying the de-convolution is currently under investigation.
 
Since each network is responsible only for a single subsection, there is no need for data exchange between processes. The training data are directly feed into the network from the memory. This avoids possible bottlenecks due to the data communication.  

\vspace{5mm}
\noindent \textbf{Inference:}
Each network can be used for the inference of the subdomain, that it is trained for. Therefore, the inference can also be done in parallel. The network receives the input at time $t$ and predicts the output at time $t+1$. For more than one time step prediction, the output at time $t+1$ can be used as the input for prediction at time $t+2$ and so on. However, the output can not be directly feed into the network, since its dimension is small. Extra data points must be received from the neighboring processes. 

For this purpose, we use MPI communication. Each processor sends the boundary data to the corresponding neighbor. To avoid the communication bottleneck, we incorporate fully parallel point-to-point communication. Each processor communicates directly to its neighbors and no central instance is used. This can be particularly important, when massively parallel machines are targeted.

\section{Numerical Performance and Accuracy Analysis}
\label{sec:results}
In this section, we numerically investigate the performance and accuracy of the proposed scheme. We showcase the learning of linearized Euler equation which is widely used in aeroacoustics simulations.    

\subsection{Test case: Linearized Euler Equation for Gaussian Pressure Pulse Investigation}
\label{test-case}
For our numerical tests, we focus on the linearized Euler equations (cf. Eq.(\ref{eq:LineulerEQ})), the linearization of the nonlinear Euler equations around a constant background. The equations to be solved allow to determine the perturbation (marked with $'$) given a known constant background (denoted with a subscript $c$) \cite{toro}
\begin{subequations} 
\label{eq:LineulerEQ}
\begin{align}
\partial_t  \rho' + \nabla \cdot \underbrace{\left( \boldsymbol{u_c} \rho' + \rho_c \boldsymbol{u'} \right)}_{:= \boldsymbol{m_u}} &= 0 \label{eq:Lineulermass} \\[1.5pt] 
 \partial_t \ \boldsymbol{u'} + \nabla \cdot \left( \boldsymbol{u_c} \boldsymbol{u'} +\frac{1}{\rho_c}p' \right) &= 0  \label{eq:Lineulermomentum} \\[1.5pt] 
 \partial_t \  p' + \nabla \cdot \left( \boldsymbol{u_c} p'	+ \gamma p_c \boldsymbol{u'} \right) &= 0 ,\label{eq:Lineulerenergy}
\end{align}
\end{subequations}

where $\rho$, $p$ and $u$ are the density, pressure and velocity, respectively. The multiplication of different perturbations are neglected, as they are insignificantly small and result in even smaller values.  \\
\textbf{Boundary condition:} At all four boundaries, outflow boundaries are considered, i.e., the pressure perturbation is set to zero, while all other quantities (density and velocity) have homogenized Neumann boundary conditions.\\ 
\textbf{Initial condition:} Initially, the fluid is at rest, the density perturbation is set to zero. The background pressure is $1~bar$ and the background density is defined to be $1~kg/m^3$. A pressure perturbation can be prescribed which corresponds to the mentioned Gaussian pulse. For this purpose, we locate a Gaussian pressure pulse in the center of our square domain at P(0.0, 0.0). The half width of the pulse is set to $0.3~m$ and the amplitude is $0.5$. The background velocity in both directions is zero.  

For comparison and to train the neural-network used in this work, the solver $Ateles$ \cite{ateles_web, roller_apes12} is used.

\subsection{Numerical Accuracy Analysis}
In this section we present the network output and compare them with validation data. The network receives the domain information, including pressure, density and velocities at time step $(t)$ and is expected to predict the domain status in the next time step. The training and validation data are produced using a numerical simulation as explained in Sec.~\ref{test-case}. We have produced in total 1500 training and validation data, by running a single simulation. We use the first 1000 time steps for the training and he remaining ones for the validation. 

\begin{figure*}[!htbp]
	\centering
	\includegraphics[width=\textwidth]{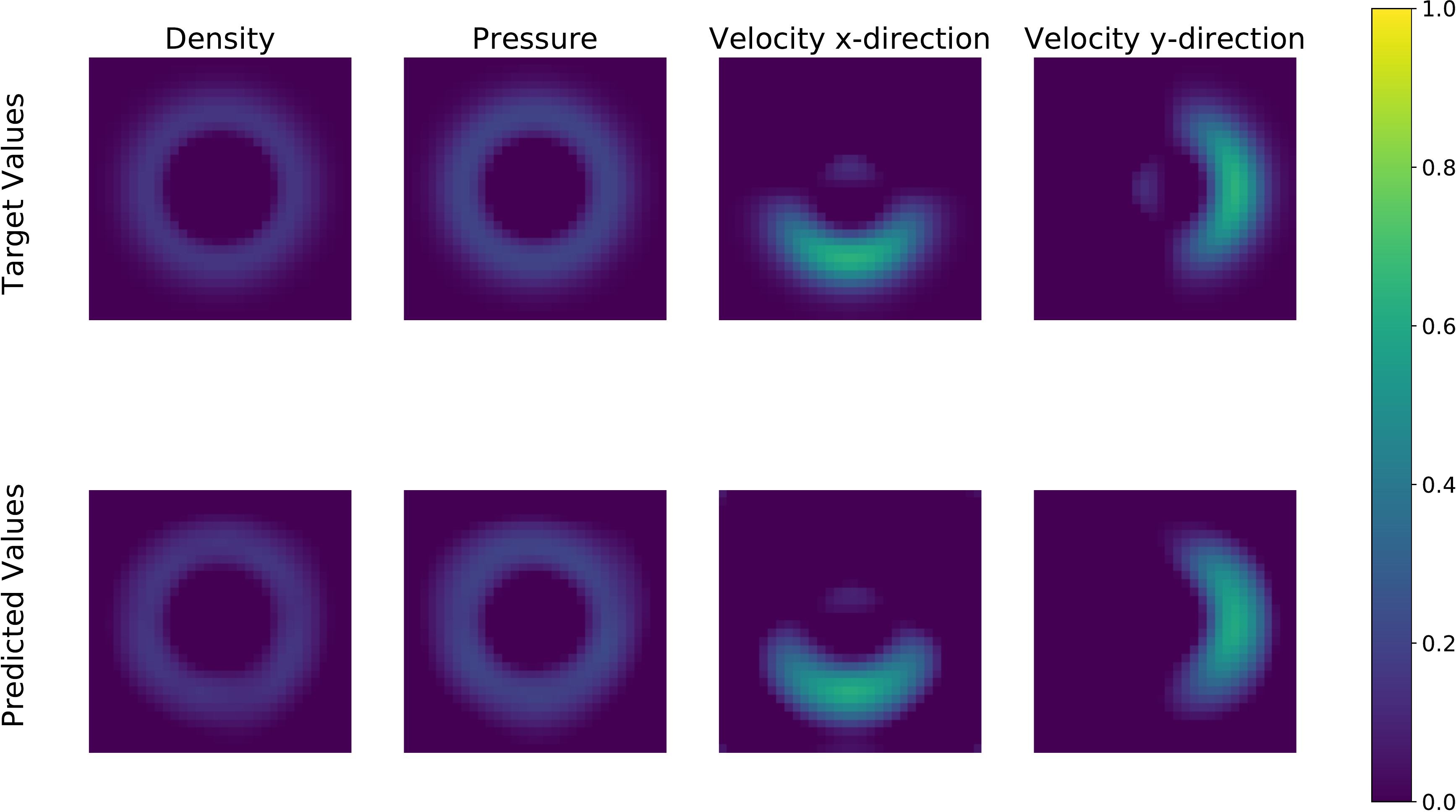}
	\caption{Comparison of the neural network output and a validation target data set for pressure, density and velocities. The input and output data are chosen randomly from the validation data set. A very good agreement between the prediction and target data can be observed.}
	\label{accurcy}
\end{figure*}

Fig.~\ref{accurcy} compares the network prediction and the target solution for the validation data sets. The comparison demonstrates a very good agreement, especially for density and pressure. There are small discrepancies in the velocities, which is probably due to the padding at the inner layers. Note that the accuracy drops after one time step prediction. This is because DNNs consisting of only CNN layers are not capable to capture the temporal connectivity completely. We always train the network with a single time step, i.e. time step $t$ is the input and $t+1$ is the output. With this setup, the network can predict a single time step accurately. However, if the output is used as a new input for the next time step prediction, the accumulative error decreases the accuracy. 

To address this problem, authors are considering incorporation of more complex layers, such as recurrent and LSTM layers. For these layers, the data must be fed into the network as time-series. This will increase the training cost, but enables the network to capture the temporal connectivity more accurately.

\subsection{Numerical Performance Analysis}
Each training data set consists of grid-type data points, 256 at each direction accumulating to a total of 65.536 points. For each point, the networks receive density, pressure, velocity in $x$- and $y$-direction. We gather the training data as a list of three-dimensional Python numpy arrays, where the $x$- and $y$-direction corresponds to the domain and the $z$-direction accounts for the various data types (density, pressure and velocities). We convert this list into standard Pytorch tensors to feed them into the network. For the parallelization, we decompose the domain into smaller subdomains as described above and use an individual network for each subdomain. 

\begin{figure}[h!]
\centering
\includegraphics[width=\linewidth]{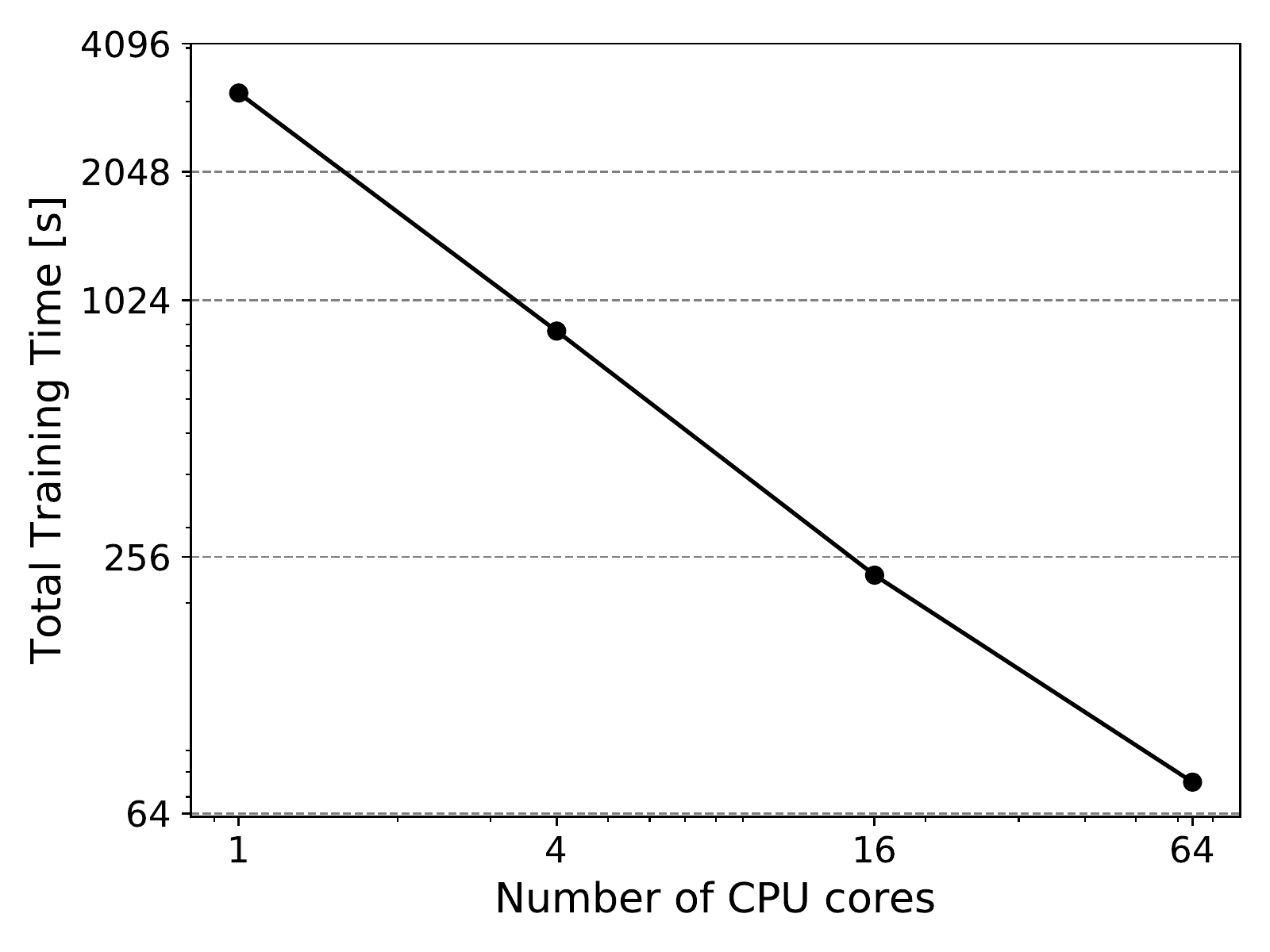}
\caption{Strong scalability analysis of the proposed parallel training scheme: The training time decreases as the number of exploited CPU cores are increased.}
\label{fig:performance}
\end{figure}

We present a strong scalability analysis for the proposed scheme up to 64 CPU cores in Fig.~\ref{fig:performance}. We observe an almost perfect strong scaling, where the training time reduces as the number of CPU cores are increased. This behavior is expected, as parallelization reduces the size of training data and thus the training time. In addition, avoiding communication during training contributes to the observed efficiency. In addition, since we do the prediction for only a single time-step, inference time is very small compared to the training.

\section{Conclusion}
\label{sec:conclusion}
A parallel scheme for training and inference of deep neural networks is introduced. The proposed method is directly applicable to simulation science for time-dependent scenarios and can be generalized for other fields. The proposed scheme is based on dividing the prediction domain into smaller subdomains and assigning each to a single MPI rank running an independent network. Since each subdomain is assigned to an individual network, no data communication is required and training data are feed to the network directly from the memory. For the inference, however, neighbouring subdomains must communicate the boundary data to preserve the spatial connectivity. To avoid any performance drop, we used fully point-to-point communication. 

The neural networks for each subdomain consists of four CNN layers with multiple channels. We showcase learning of linearized Euler equations, with an application for aeroacoustics simulations. Accordingly, the input data has four channels, corresponding to pressure, density and velocities in $x$- and $y$-direction. To preserve the output dimension for each layer, we incorporated padding. The inputs of each neuron are mapped to the output using a leaky ReLU activation function. The MAPE loss function is used along with the ADAM optimization method to calculate the network's weights. 

A first numerical accuracy analysis demonstrated a very good agreement between network prediction and simulation data. In addition, the strong scalability analysis showed a very good reduction in training time by increasing the number of CPU cores. The predictions accuracy can be further improved by incorporating more complex layers such as LSTM and enhancing training data to account for the temporal connectivity. This topic is currently under investigation by the authors.

\section*{Acknowledgment}
We thank the Deutsche Forschungsgemeinschaft (DFG, German Research Foundation) for supporting this work by funding - EXC2075 – 390740016 under Germany's Excellence Strategy. We acknowledge the support by the Stuttgart Center for Simulation Science (SimTech).
The current work as also financially supported by the priority program 1648--Software for Exascale Computing 214 (www.sppexa.de) of the Deutsche Forschungsgemeinschaft.

\bibliographystyle{unsrt}
\bibliography{paper}

\end{document}